\documentclass{svjour3}                     
\smartqed  
\usepackage[T1]{fontenc}
\usepackage[utf8]{inputenc}
\usepackage[english]{babel}

\usepackage{amssymb,amsfonts,amsmath}
\usepackage{braket}
\usepackage{graphicx}
\usepackage{hyperref}
\usepackage{enumerate}
\usepackage{multirow}
\usepackage{booktabs}
\usepackage{caption}
\usepackage{float}

\newcommand{\Tr}[0]{\operatorname{Tr}}

\newcommand{\TrSet}[0]{\mathcal{S}_{\mathrm{tr}}}
\newcommand{\TsSet}[0]{\mathcal{S}_{\mathrm{ts}}}

\newcommand{\Real}[0]{\mathbb{R}}
\newcommand{\argmin}[0]{\operatorname*{argmin}}
\newcommand{\TrQSet}[0]{\mathcal{S}_{\mathrm{tr}}^\mathrm{q}}

\newcommand{\dtr}[0]{d_\mathrm{tr}}

\begin{document}

\title{Quantum-inspired Minimum Distance Classification in Biomedical Context 
}

\titlerunning{Quantum-inspired Minimum Distance Classification in Biomedical Context }        

\author{Giuseppe Sergioli, Giorgio Russo, Enrica Santucci, Alessandro Stefano, Sebastiano Emanuele Torrisi, Stefano Palmucci, Carlo Vancheri, Roberto Giuntini
}

\authorrunning{ G Sergioli, G Russo, E Santucci, A Stefano, S Torrisi, S Palmucci, C Vanchieri, R Giuntini} 

\institute{Giuseppe Sergioli \at
              University of Cagliari,
             Via Is Mirrionis 1, I-09123 Cagliari, Italy\\
              \email{giuseppe.sergioli@gmail.com}\\
Giorgio Russo \at
              Institute of Molecular Bioimaging and Physiology, National Research Council (IBFM-CNR),
             Contrada Pietropollastra-Pisciotta, I-90015 Cefal\`u, Italy\\
              \email{giorgio.russo@ibfm.cnr.it}\\
Enrica Santucci \at
              University of Cagliari,
             Via Is Mirrionis 1, I-09123 Cagliari, Italy\\
              \email{enrica.santucci@gmail.com}                      \\
Alessandro Stefano \at
              Institute of Molecular Bioimaging and Physiology, National Research Council (IBFM-CNR),
             Contrada Pietropollastra-Pisciotta, I-90015 Cefal\`u, Italy\\
              \email{alessandro.stefano@ibfm.cnr.it}
\\
Sebastiano Emanuele Torrisi \at
              Regional Referral Centre for Rare Lung Diseases, A.O.U. Policlinico-Vittorio Emanuele, University of Catania, 
              Catania, Italy\\
              \email{torrisiseby@hotmail.it}\\
Stefano Palmucci \at
              Radiodiagnostics and Oncological Radiotherapy Unit, University Hospital "Policlinico-Vittorio Emanuele", 
              Catania, Italy\\
              \email{spalmucci@sirm.org}\\
Carlo Vancheri \at
              Regional Referral Centre for Rare Lung Diseases, A.O.U. Policlinico-Vittorio Emanuele, University of Catania, 
              Catania, Italy\\
              \email{vancheri@unict.it}\\
Roberto Giuntini \at
              University of Cagliari,
             Via Is Mirrionis 1, I-09123 Cagliari, Italy\\
              \email{giuntini@unica.it}\\}

\date{Received: date / Accepted: date}

\maketitle
\begin{abstract}
We face the problem of pattern classification by proposing a quantum-inspired version of the widely used minimum distance classifier (\emph{i.e.} the Nearest Mean Classifier (NMC)) already introduced in \cite{sergioli2016,S,SaSe,Entropy} and by applying this quantum-inspired classifier in a biomedical context. In particular, we show and compare the NMC and our quantum model performance to solve a problem related to classify the probability of survival for patients affected by idiopathic pulmonary fibrosis (IPF).
\end{abstract}

\section{Introduction}
Quantum mechanics is a probabilistic theory that turns out to be particularly suitable to describe different kinds of stochastic processes, that - in principle - can also include non-microscopic domains.
As some example, in recent years quantum formalism has been exploited in non standard contexts such as game theory, economic processes, cognitive sciences and so on \cite{1,2,5,14,hla,qicm}.

By this perspective, another non-standard application of quantum theory is devoted to apply it for solving classification problems. The basic idea is to represent classical patterns in terms of quantum objects, with the aim to boost the computational efficiency of the classification algorithms. In the last few years many efforts have been made to apply the quantum formalism to signal processing \cite{7} and pattern recognition \cite{8,9}. Exhaustive surveys concerning the applications of quantum computing in computational intelligence and machine learning are provided in \cite{10,11}. Even if these approaches suggest possible computational advantages of
this sort \cite{12,13}, what we have proposed in \cite{sergioli2016,S,SaSe,Entropy} is based on a different approach that consists in using quantum formalism in order to reach remarkable benefits in the classical context. What we have provided is a model that allows to process any kind of classical dataset in a supervised system by $i)$ translating each element of the dataset (pattern) into a density operator (that is the usual mathematical tool to formally describe a quantum state) that will be called \emph{density pattern}; $ii)$ defining, for any class of density patterns, a \emph{quantum centroid} that is an object free of any counterpart in the initial classical dataset; $iii)$ using the standard minimum distance procedure to classify an unlabeled density pattern; $iv)$ decoding the result of the classification process in the classical pattern space.
In this way, by exploiting the expressive power of the quantum formalism, it is possible to reach remarkable advantages in terms of classification process accuracy.
In this regard, we have shown a comparison between the standard \emph{Nearest Mean Classifier} (NMC) and its quantum version (named \emph{Quantum Nearest Mean Classifier} (QNMC)), exhibiting meaningful advantages of our proposed model in its application on different datasets. In particular, the model has been tested on artificial and real datasets commonly downloadable by standard machine learning repositories.
In the present work we propose a particular application of the model to a real dataset (IPF dataset) that is obtained form a group of 126 patients. IPF is a disease characterized by the development of fibrotic areas within the parenchyma of lungs causing a progressive reduction of the respiratory function. The prognosis of IPF patient is very poor with a median survival of 3-5 years from diagnosis; the dataset includes baseline variables with an established relation to patient's survival. In this paper we refer to the IPF dataset to compare the performances of two different variants of the QNMC not only with the NMC but also with other well known standard classifiers (the \emph{Linear Discriminant Analyisis} (LDA) and the \emph{Quadratic Discriminant Analysis} (QDA)).

The paper is organized as follows: in the first section we briefly describe the formal structure of both the Nearest Mean Classifier and its quantum-inspired version (the QNMC). In the second section we briefly summarize some interesting results previously obtained in \cite{sergioli2016,S,SaSe,Entropy} by comparing the NMC and the QNMC on different datasets, and showing the advantages of the QNMC in terms of pattern classification accuracy. In the third section we first introduce an alternative encoding from the real vector (pattern) space to the density operator space that turns out to be particularly beneficial for the rest of the paper. Secondly, we introduce the IPF dataset and we provide a detailed description of the dataset features. Finally, we show and discuss the promising results arising by the application of two different QNMC variants on the IPF dataset, showing an improvement of the accuracy with respect to some standard classifiers, \emph{i.e.} the NMC, the LDA and the QDA. The last section of the paper is devoted to propose possible developments and different strategies we will take into account in future works in order to provide a further improvement in terms of classification accuracy in biomedical contexts.

\section{Classical and quantum version of the nearest mean classifier}
\label{sec:NMC}

In this section we briefly describe the quantum version of the standard nearest mean classification,  which is an instance of supervised learning, \emph{i.e.} learning from a training dataset of correctly labeled objects. In the classical domain each object is characterized by its features; hence, a $d$-feature object is naturally represented by a $d$-dimensional real vector $\vec x = \left[x^1, \ldots, x^d \right]\in \Real^d.$\footnote {For the sake of clarity regarding the indexes, we accord to use superscript index to indicate the different components of the vector and subscript to indicate different vectors.}
 Formally, a pattern can be represented as a pair $(\vec x_i, \lambda_i)$, where $\vec x_i$ is the $d$-dimensional vector associated to the object and $\lambda_i$ is the label that refers to the class which the object belongs to. We can simply consider a class as a set of objects and, for our aim, we confine ourselves to the special (but very common) case where each object belongs to one and only one class of objects. Let $\Lambda=\{\lambda_1, \ldots, \lambda_N\}$ be the set of labels corresponding to the respective classes. The goal of the classification process is to design a \emph{classifier} that attributes (in the most accurate way) a label (class) to any unlabled object. In supervised learning, such a classifier is obtained by getting information from the \emph{training set} $\TrSet$, \emph{i.e.} a set of correctly labeled objects. Formally: 
$$\TrSet = \left\{ (\vec x_1, \lambda_1), \ldots , (\vec x_M, \lambda_M) \right\},$$
where $\vec x_i\in\mathbb R^d$ and $\lambda_i$ is the label associated to its class.  

Generally, we will deal with $n$ possible different classes (\emph{i.e.} $N=n$) and,
given a training dataset $\TrSet = \left\{ (\vec x_1, \lambda_1), \ldots , (\vec x_M, \lambda_M) \right\}$, we can define the $j$-th class $\TrSet^{j}$, which represents the set of the training patterns belonging to the class labeled by $\lambda_j$, in the following way:
$$\TrSet^{j}=\{(\vec x_i, \lambda_i)\in\TrSet : \lambda_i=\lambda_j\}.$$
Finally, by $M_j$ we will denote the number of elements of $\TrSet^j$.

One of the simplest classification method in pattern recognition is the so called \textit{Nearest Mean Classifier}. The NMC algorithm consists in the following steps:
\begin{enumerate}
  \item Training: one has to compute the \textit{centroid} for each class, that is:
  \begin{equation}\label{eq:ccentroid}
  \vec \mu_j = \frac{1}{M_j} \sum_{i\in\{m\in M:\lambda_m=\lambda_j\}} \vec x_i 
  \end{equation}
 \item Classification: the associated classifier is a function $Cl:\mathbb R^d\to\Lambda$ such that $\forall \vec x\in \mathbb R^d$:
$$Cl(\vec x)=
\lambda_j  \,\,\,\,\text{if} \,\,\,\,d(\vec x,\vec \mu_j)\leq d(\vec x,\vec \mu_k) \,\, \forall k\neq j.$$
where $d(\vec x, \vec y)=|\vec x - \vec y|$ is the Euclidean distance. 
\end{enumerate}
Intiutively, the classifier associates to a $d$-feature object $\vec x$ the label of the closest centroid.

%
In order to evaluate the NMC performance, one introduces another set of patterns (called \textit{test set}) that does not belong to the training set \cite{DuHa}. Formally, the test set is a set $\TsSet = \left\{ \{\vec y_{1}, \beta_{1}\}, \ldots , \{\vec y_{M'}, \beta_{M'}\} \right\}$, such that $\TrSet\cap\TsSet=\emptyset$, where $M'_j$ is the number of the test patterns belonging to the $j$-th class.

Then, by applying the NMC to the test set, it is possible to evaluate the semi-supervised classifier performance by considering the accuracy (ACC) of the classification process as the ratio between the number of all the correctly classified test patterns and the cardinality of the test set.\footnote{We recall that the classification accuracy is defined as ACC $= 1 - $ERR, where ERR is the classification error. Consequently, it is possible to study the performance of a given classification method by means of accuracy or error likewise.}

Let us notice that the values of such quantities are obviously related to the training/test datasets; as a natural consequence, also the classifier performance is strictly dataset-dependent.

\bigskip

In order to provide a quantum counterpart of the NMC (we say \emph{Quantum Nearest Mean Classifier} (QNMC)) we need to fulfill the  following steps:
\begin{enumerate}
  \item for each pattern, one has to provide a suitable encoding into a quantum object (\emph{i.e.} a density operator) that we call \textit{density pattern};
  \item for each class of density patterns, one has to define the quantum conterpart of the classical centroid, that we call \textit{quantum centroid};
  \item one has to provide a suitable definition of \textit{quantum distance} between density patterns, that plays a similar role as the Euclidean distance for the NMC.
\end{enumerate}

Even though there are infinite many ways to encode a real vector into a density pattern (and the convenience of using one instead of others could be strictly dataset-dependent), in \cite{S} we have propose the following encoding, that we call \emph{stereographic encoding} (SE).

First, let us recall the notion of \emph{stereographic projection} as follows. 
Let $\vec {\tilde x} = \left[\tilde x^1, \ldots, \tilde x^{d+1} \right]$ be an arbitrary $(d+1)$-feature object of $\mathbb R^{d+1}$. The stereographic projection $SP$ is a map $SP:\mathbb R^{d+1}\to\mathbb R^d$ such that:
$$\vec x=SP(\tilde{x}^1,\tilde{x}^2,...,\tilde{x}^{d+1})=\Big (\frac{\tilde{x}^1}{1-\tilde{x}^{d+1}},...,\frac{\tilde{x}^d}{1-\tilde{x}^{d+1}}\Big ).$$
Analogously, let $\vec x = \left[x^1, \ldots, x^d \right]$ be an arbitrary $d$-feature object of $\mathbb R^d$; 
the inverse of the stereographic projection $SP^{-1}$, is a map $SP^{-1}:\mathbb R^{d}\to\mathbb\mathbb R^{d+1}$ such that:
\begin{equation}\label{eq:strproj}
  \vec{ \tilde{x}} = SP^{-1}(\vec x) = \frac{1}{\sum_{i=1}^d (x^i)^2+1} \left[2x^1, \ldots, 2x^d, \sum_{i=1}^d (x^i)^2 - 1 \right],
\end{equation}

where $\frac{1}{\sum_{i=1}^d (x^i)^2+1}$ is a normalization factor.
\begin{definition}[Density pattern by SE]
\label{def:dp}

 The density pattern $\rho_{\vec x}$ associated to the $d$-feature object $\vec x\in\mathbb R^d$ is defined as:
\begin{equation}
 \label{eq:dp}
 \rho_{\vec x} \doteq \vec{ \tilde{x}}^t \vec{ \tilde{x}}.
\end{equation}
\end{definition}

Clearly, every density pattern is a quantum pure state, \emph{i.e.} $\rho_{\vec x}^2 = \rho_{\vec x}$.
Therefore, the SE allows to encode any real vector $\vec x\in\mathbb R^d$ into a density operator $\rho_{\vec x}$.  On this basis, we define the quantum training dataset 

$$\TrQSet = \left\{ \{\rho_{\vec x_1}, \lambda_1\}, \ldots , \{\rho_{\vec x_M}, \lambda_M\} \right\}$$
as the set of all the density patterns obtained by encoding all the elements of $\TrSet$.

This fact allows us to introduce the quantum versions of the standard centroid given in Eq.~\eqref{eq:ccentroid}, as following.
\begin{definition}[Quantum centroids]
\label{def:qcentroid}
Let $\TrQSet = \left\{ \{\rho_{\vec x_1}, \alpha_1\}, \ldots , \{\rho_{\vec x_M}, \alpha_M\} \right\}$ be a quantum training dataset of density patterns.
The quantum centroids for the positive and negative class are respectively given by:
\begin{equation}
\label{eq:qcentroid}
\rho_j = \frac{1}{M_j} \sum_{i\in\{m\in M:\lambda=\lambda_j\}} \rho_{\vec x_i}.
\end{equation}
\end{definition}
Notice that the quantum centroids are now mixed states and they are not generally obtained by the encoding of the respective classical centroids $\vec \mu_j$.
Accordingly, the definition of the quantum centroid leads to a new object that does not have any classical counterpart.

As a suitable definition of distance between density patterns, we recall the well known distance between quantum  states that is commonly used in quantum computation (see, \emph{e.g.}~\cite{nielsenbook}).
\begin{definition}[Trace distance]
\label{def:trdist}
Let $\rho$ and $\sigma$ be two quantum density operators belonging to the same Hilbert space. The trace distance ($\dtr$) between $\rho$ and $\sigma$ is given by:
\begin{equation}\label{eq:trdist}
  \dtr(\rho,\sigma) = \frac{1}{2} \Tr|\rho - \sigma|,
\end{equation}
where $|A| = \sqrt{A^\dag A}$.
\end{definition}
Notice that the trace distance is a metric; hence, it satisfies: \emph{i)} $\dtr(\rho,\sigma) \geq 0$ with equality iff $\rho=\sigma$ (positivity), \emph{ii)}   $\dtr(\rho,\sigma) =   \dtr(\sigma,\rho)$ (symmetry) and \emph{iii)} $\dtr(\rho,\omega)+\dtr(\omega,\sigma) \geq \dtr(\rho,\sigma)$ (triangle inequality).

We have introduced all the necessary ingredients to describe into detail the QNMC process, which, similarly to the classical case, consists in the following steps:
\begin{itemize}
  \item obtaining the quantum training dataset $\TrQSet$ by applying the encoding  given in Definition~\ref{def:dp} to each pattern of the classical training set $\TrSet$;
  \item calculating the quantum centroids $\rho_j$ according to Definition~\ref{def:qcentroid};
  \item classifying an arbitrary pattern $\vec x$ accordingly with the minimization problem:
the quantum classifier is a function $QCl:\mathbb R^d\to\Lambda$ such that $\forall \vec x\in \mathbb R^d$:
$$QCl(\rho_{\vec x})=
\lambda_j \,\,\,\,\text{if} \,\,\dtr(\rho_{\vec x},\rho_{j})\leq \dtr(\rho_{\vec x},\rho_{i}) \,\,\,\, \forall i\neq j.$$

\end{itemize}

\subsection{Experimental results}
\label{sec:exp}

In what follows we summarize some preliminary result obtained by comparing the performances of NMC and the QNMC on different (artificial and real) different datasets. In particular, we consider
three artificial (two-feature) datasets  (Moon, Banana, and Gaussian) and four real (many-feature) datasets (Diabetes, Cancer, Liver and Ionosphere) extracted from the UC Irvine Machine Learning Repository.

In our experiment, we follow the standard methodology of randomly splitting each dataset in training and test datasets with $\%80$ and $\%20$ of the total patterns, respectively. Moreover, in order to obtain statistical significance results, we carry out $100$ experiments for each dataset, where the splitting is randomly taken each time.

We summarize our results in the Table \ref{Tab}.
\begin{table}[h]
\tiny{
\begin{center}
\begin{tabular}{cccccccc}
\midrule
Dataset    & $\#\TrSet$/$\#\TsSet$& ACC  (NMC)  &   ACC (QNMC) \\
\midrule
\vspace{0.1cm}
    Banana &  $4240/1060$($2$)    & $55.0\pm 1.8$ & $\mathbf{71.0\pm 1.2}$ 
\\
\vspace{0.1cm}
   Gaussian &  $160/40$($2$)    & $55.5\pm 7.7$ &  $\mathbf{76.2\pm 5.6}$
\\
\vspace{0.1cm}
    Moon &  $160/40$($2$)    & 77.9$\pm 5.7$ &  $\mathbf{88.9\pm 4.4}$
\\
\vspace{0.1cm}
Diabetes    &  $614/154$($8$)    &63.4 $\pm 3.9$ &  $\mathbf{68.7\pm 3.2}$ 
\\
\vspace{0.1cm}
Cancer    &  $546/137$ ($10$)    & $\mathbf{96.4\pm 1.4}$ &  $93.7\pm 1.9$
\\
\vspace{0.1cm}
    Liver&  $463/116$($10$)    & $53.8\pm 4.2$ & $\mathbf{59.6\pm 4.2}$ 
\\
\vspace{0.1cm}
Ionosphere    &  $280/71$($34$) & $72.9\pm 4.5$ &  $\mathbf{83.7\pm 4.3}$
\\
%

\bottomrule
\end{tabular}
\end{center}
\caption{Average results for NMC and QNMC classifiers (in $\%$) and their standard deviations. $\#\TrSet$ cardinality of the training dataset; $\#\TsSet=$ cardinality of the test set; $(d)=$ number of features of each element of its respective dataset.}
\label{Tab}
}
\end{table}

Let us notice that the ACC of the QNMC is significantly greater than the ACC of the NMC for all the datsets, except for the Cancer dataset.
In particular, this improvement is even greater for the $2$-feature datasets.

Further, let us notice that a key difference between NMC and QNMC regards the invariance under rescaling. Let us suppose that each pattern of the training and test sets is multiplied by the same \textit{rescaling factor} $t$, \emph{i.e.} $\vec{ x}_m \mapsto t \vec{ x}_m$ and $\vec{ y}_{m'} \mapsto t \vec{ y}_{m'}$ for any $m$ and $m'$. Then, the (classical) centroids change according to $\vec{\mu}_j \mapsto t \vec{\mu}_j$ and the classification problem of each pattern of the rescaled test set becomes
  \begin{equation}\label{eq:rescaledcclassify}
  \argmin_i d(t \vec y_{m'}, t \vec \mu_i) = t  \argmin_i d(\vec y_{m'},\vec \mu_i),
  \end{equation}
which has the same solution of the unrescaled problem, \emph{i.e.} $t=1$.

\begin{figure}[h]
      \centering
  \includegraphics[width=\textwidth]{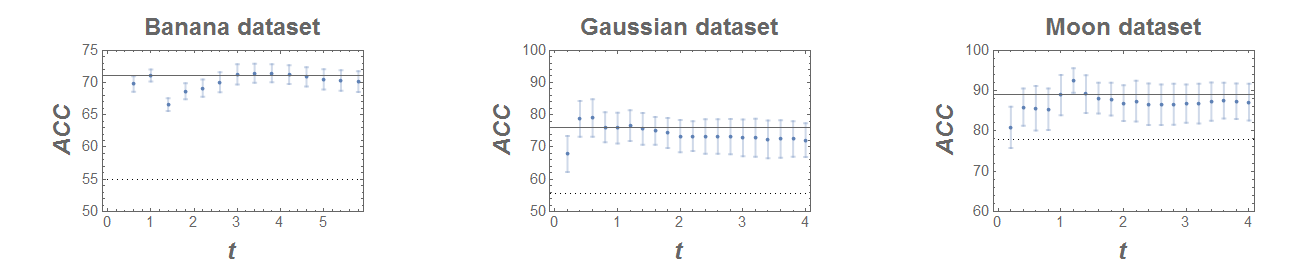}
  \\
  \caption{Accuracy vs rescaling factor for Banana (left), Gaussian (medium) and Moon (right) datasets. For a fixed $t$, each point corresponds to an average of the accuracy under $20$ experiments and its standard deviation is represented by the corresponding error bar. In each plot, we fix the $y$-axis to the case $t=1$ (no rescaling). The dotted line represents the average accuracy reached by the NMC.
}
\label{rescaling}
\end{figure}

On the contrary, the QNMC turns out to be not invariant under rescaling.
Far from being a disadvantage, this allows us to introduce a ``free'' parameter, \emph{i.e.} the \emph{rescaling factor}, that could be useful to get a further improvement of the classification performance as it is shown in Fig.~\ref{rescaling} for the $2$-feature datasets. The pictures in Fig.~\ref{rescaling} represent the experimental results where we repeated the same experiments described above by rescaling (within a small range) the coordinates of the initial dataset. The picture of the Figure 1 shows that for each dataset  there is an interval ($I_t$) of the rescaling factor $t$ such that for any $t\in I_t$ the average values of the accuracy are slightly greater than the accuracy values of the respective unrescaled cases. 

The method that we have introduced in this section allows us to get a relevant improvements of the standard NMC when we have an \emph{a priori} knowledge about the distribution of the dataset we have to deal with. Indeed, if we need to classify an unknown pattern, looking at the distribution of the training dataset, we can guess \emph{a priori} if: \emph{i)} for that kind of distribution the QNMC performs better than the NMC and \emph{ii)} what is the suitable rescaling has to be applied to the original dataset in order to get a further improvement of the accuracy.

\section{Applying the QNMC on the IPF dataset}

As mentioned at the beginning of the previous section, there generally exist different ways to encode a $d$-dimensional feature vector into a density operator \cite{8}. Indeed, finding the ``best'' encoding of real vectors into quantum states (\emph{i.e.} outperforming all the possible encodings for any dataset) is still an open and intricate problem. This fact is not so surprising because, on the other hand, in pattern recognition is not possible to establish an absolute and \emph{a priori} superiority of a given classification method with respect to the other ones, and the reason is that each dataset has unique and specific characteristics (according to the well known \emph{No Free Lunch} Theorem \cite{DuHa}).
 
Hereby, we like to introduce a new encoding, we call \emph{informative encoding} (IE), already used in two previous works \cite{SaSe,Entropy}. According to recent debates on quantum machine learning \cite{8}, in order to avoid loss of information it is crucial that, in the transition from the classical to the quantum space, the quantum state keeps information on the original vector norm.\\

\noindent Let $\vec x = (x^1, \ldots, x^d)\in \mathbb{R}^d$ be a $d$-dimensional vector.
\begin{enumerate}
\item We map the vector $\vec x \in \mathbb{R}^d$ into a vector $\vec x' \in \mathbb{R}^{d+1}$, whose first $d$ features are the components of the vector $\vec x$ and the $(d+1)$-th feature is the norm of $\vec x$. Formally:
\begin{equation}
\vec x = (x^1, \ldots, x^d)\ \mapsto\ \vec x' = (x^1, \ldots, x^d, |\vec x|).
\end{equation}
\item We obtain the vector $\vec x''$ by dividing the first $d$ components of the vector $\vec x'$ for $|\vec x|$: 
\begin{equation}
\label{x''}
\vec x' \ \mapsto\ \vec x'' = \Big (\frac{x^1}{|\vec x|}, \ldots, \frac{x^d}{|\vec x|}, |\vec x|\Big ).
\end{equation}
\item We compute the norm of the vector $\vec x''$, \emph{i.e.} $|\vec x''| = \sqrt{|\vec x|^2 + 1}$ and we map the vector $\vec x''$ into the normalized vector $\vec x'''$ as follows: 
\begin{equation}\label{x'''}
\vec x'' \ \mapsto\ \vec x''' = \frac{\vec x''}{|\vec x''|}= \Big (\frac{x^1}{|\vec x|\sqrt{|\vec x|^2 + 1}}, \ldots, \frac{x^d}{|\vec x|\sqrt{|\vec x|^2 + 1}}, \frac{|\vec x|}{\sqrt{|\vec x|^2 + 1}}\Big ).
\end{equation}
\end{enumerate}
Now, similarly to Definition \ref{def:dp}, we end up with the following definition.
\begin{definition}[Density pattern by IE]

\begin{equation}
 \label{eq:dp}
\rho_{\vec x} \doteq \vec x'''\cdot(\vec x''')^\dagger,
\end{equation}
where the vector $\vec x'''$ is given by Eq. (\ref{x'''}).
\end{definition}

Hence, this encoding maps real $d$-dimensional vectors $\vec x$ into a $(d+1)$-dimensional pure state $\rho_{\vec x}$. In this way, we obtain an encoding that takes into account the information about the initial real vector norm and, at the same time, allows to easily encode arbitrary real $d$-dimensional vectors.\\
\noindent As we have seen in the previous section, the QNMC is the quantum replacement of the standard NMC that is one of the more basic standard classifiers. Other well known standard models that will be taken into account in the following are the so called \emph{Linear Discriminant Analysis} (LDA) and \emph{Quadratic Discriminant Analysis} (QDA) classifiers \cite{DuHa}. In particular, they belong to the set of minimum distance classifiers and the goal consists in classifying patterns by using a distance measure which involves not only the centroids of the classes but also the class distribution (by means of the \emph{covariance matrix} \cite{DuHa}). The difference between them is the following: \emph{i)} in the LDA case the distance measure depends on the average covariance matrix (over all the covariance matrices related to each class) and the discriminant function (\emph{i.e.} the surface which separates classes in the optimal way) is linear; \emph{ii)} in the QDA case, the distance measure depends on all the covariance matrices simultaneously and the discriminant function is quadratic.
In what follows, we compare different variants of the QNMC with the mentioned classifiers (NMC, LDA, QDA) by referring to a very special real dataset obtained from a biomedical context. \footnote{The dataset is downloadable from

 \emph{http://people.unica.it/giuseppesergioli/files/2018/02/IPFDataset.xlsx}}

\subsection{The IPF dataset}
In details, the idiopathic pulmonary fibrosis (IPF) dataset includes a group of 126 consecutive patients (the patterns) retrospectively extracted from databases of the Regional Referral Centre for Interstitial and Rare lung diseases of Catania. These patients are divided in three different classes (with different cardinality), where each class corresponds to a different degree of survival (that is named GAP stage). All patients were required to have received a Multidisciplinary team diagnosis of IPF according to 2011 American Thoracic Society (ATS)/European Respiratory Society (ERS)/Japanese Respiratory Society (JRS)/Latin American Thoracic Association (ALAT) IPF guidelines \cite{Sebastiano1}. A minimum follow-up time of three years from diagnosis was also required in order to assess survival. For this reason, only patients diagnosed between July 2010 and December 2014 were considered. The dataset includes a series of baseline variables (the features) with an established relation to survival (the classes, where three different survival “degrees” are considered) \cite{Sebastiano2,Sebastiano3}.

The dataset is organized in the following way: the patterns are numered in the column  A (we also indicate in the column B the dates of birth of each patient). We distinguish between two different blocks of features; the first block (from column C to column I, highlighted in light grey) contains features that allow to perfectly classify each patient; indeed, by using the features introduced in the columns C ... I, it is possible to exactly evaluate the “GAP stage” of each patient (each feature adds a score to the calculation of the GAP stage). Indeed, the features introduced from C to I are all it takes in order to assign to each patient the class to which he belongs to; in other words, these features are useful to have an \emph{a priori} classification of each patient. The second block of features are introduced (in light green) from column J to column U; even if these  features should allow to classify the patients, anyway - unlike the first block - there is not a systematic method to classify each patient by involving this set of features only. 
Finally, the column W contains the labels associate to each different class (the column V is only used as a support to calculate W). The rest of the paper will be devoted to use the introduced quantum-inspired algorithm to classify the IPF dataset, only involving the second block of features. But before let us briefly provide a medical description of the meaning of each feature.

Regarding the first block, the feature “Forced Vital Capacity” (FVC) represents the amount of air which can be forcibly exhaled from the lungs after taking the deepest breath possible \cite{Sebastiano4}. This value, measured with a spirometer, was reported in the dataset as percent of predicted value (FVC\%), resulting from the comparison between a list of normal reference values and the measured ones \cite{Sebastiano4}. In the context of IPF, both baseline FVC\% value and its change over the time, represent strong predictors of mortality \cite{Sebastiano5,Sebastiano6}. The feature  “Diffusing Capacity for Carbon Monoxide” (DLCO), measures the ability of the lungs to transfer gas from inhaled air to the red blood cells in pulmonary capillaries \cite{Sebastiano7}. As in the case of FVC, also DLCO is expressed as percent of predicted value. Interestingly in IPF, DLCO is frequently reduced since early stages of the disease, making this variable more sensitive than FVC to assess interstitial lung damage \cite{Sebastiano8}. Another feature collected which significantly impacts on survival, as in IPF as in other diseases, is the “Age at first diagnosis” \cite{Sebastiano1}. Dataset also included the variable “Sex”. Incidence and prevalence of IPF are higher in males than in females with a ratio ranging from 1.6:1 to 2:1.  Moreover, male sex was demonstrated to be related with a worse prognosis \cite{Sebastiano1,Sebastiano10}. All of these four features were recently included in a single multidimensional index, known as GAP (gender [G], age [A] and lung physiology variables [P]). This index assigns a point to each variable in order to obtain a single value, in the dataset “GAP point”, which resumes the weight of each variable. Points raging from 0 to 3, 4-5 and 6-8 compose respectively “GAP stage 1, 2 and 3” \cite{Sebastiano11}, that we consider as the label of our dataset. Simply speaking, the columns from F to I indicate the contribute in the calculation of the GAP stage provided by the features “Sex”, “FVC”, “DLCO” and “Age”, respectively. Regarding the second block of features, Oxygen saturation (SpO2 \%) reflects blood oxygenation, and heart rate were indirectly measured with a pulse oximeter. Reduced levels of SpO2, which are frequently associated with high levels of heart rate, are usually related to a worse survival \cite{Sebastiano1}. Information regarding smoking habit was also collected and reported as follows: never smoker =0, ex smoker =1 and current smoker=2. Dataset included also a description of high resolution computed tomography (HRCT) features which, according to 2011 IPF guidelines, describe three scenarios: “definite UIP”, “possible UIP” and “inconsistent with UIP” \cite{Sebastiano1}. Recent studies demonstrated that also this evaluation at baseline is related with prognosis \cite{Sebastiano12}. Other variables regarding information on lung transplantation, duration of follow-up (days), status at the end of follow-up (alive $=0$ or died $=1$), confirmation of diagnosis through biopsy and family history of the Interstitial Lung Diesease (ILD) were also included in the dataset.\\

\subsection{Applying  the QNMC to the IPF dataset}

It is natural to believe how each feature described above has not the same impact in the evaluation of the GAP stage (i.e. in the classification process). As an example (confining to the second block of features only), it is possible to say that “Sex” and “Oxygen Saturation” have more impact in the classification process with respect to the rest of the considered features. In general, it is possible to  recognise for each feature a different impact in the classification process.

\noindent On this basis, let us stress that the key difference between NMC and QNMC regards the invariance under rescaling \cite{sergioli2016,S,Entropy}. Indeed, we have shown in the previous works that, conversely to the standard NMC, the QNMC turns out to be not invariant under rescaling, \emph{i.e.} if we multiply each dataset pattern for a real factor. Far from being a disadvantage, this allows to introduce a “free” parameter that could be useful to get a further improvement of the classification performance.\\
\noindent In order to take into suitable account both the different incidence of each dataset feature in the classification process and the non-invariance of the QNMC under rescaling, the strategy we adopt is to assign for each feature a rescaling factor that is proportional to its degree of incidence. Differently from the previous section, where all the dataset features were multiplied for the same rescaling real factor, here we multiply each feature for a different weight, in accord with the incidence of each feature in the evaluation of the GAP stage (that is related to the degree of survival). Consequently, the rescaled dataset becomes:
\begin{equation}\label{rescdataset}
\mathcal{S}^{(r)} = \TrSet^{(r)} \cup \TsSet^{(r)}
\end{equation}
where
\begin{align}
\TrSet^{(r)} &= \left\{ (\gamma_1 \vec x_1, \lambda_1), \ldots , (\gamma_M \vec x_M, \lambda_M) \right\},\quad \gamma_i\in \mathbb{R},\ i,=1, \ldots, M,\nonumber\\
\TsSet^{(r)} &= \left\{ \{\delta_1 \vec y_{1}, \beta_{1}\}, \ldots , \{\delta_{M'} \vec y_{M'}, \beta_{M'}\} \right\}, \quad \delta_{j}\in \mathbb{R},\ j=1, \ldots, M'. \nonumber
\end{align}
Finally, the quantum version $\mathcal{S}^{\text{q}(r)} = \mathcal{S}_{tr}^{\text{q}(r)}\cup \mathcal{S}_{ts}^{\text{q}(r)}$ of the rescaled dataset is obtained by putting $\rho_{\vec x_i},\ \rho_{\vec y_j}$ in place of $\vec{x}_i,\ \vec{y}_j$ in Equation (\ref{rescdataset}).

\begin{table}[h]
\tiny{
\begin{center}
\begin{tabular}{cc}
\midrule
Classifier & Total Error \\
\midrule
\vspace{0.1cm}
QNMC (SE) & 0.455 $\pm$ 0.093 \\
\vspace{0.1cm}
QNMC (IE) & 0.378 $\pm$ 0.092 \\
\vspace{0.1cm}
QNMC (IE) Resc 1 & 0.334 $\pm$ 0.097 \\
\vspace{0.1cm}
QNMC (IE) Resc 2 & 0.341 $\pm$ 0.071 \\
\vspace{0.1cm}
QNMC (IE) Resc 3 & 0.344 $\pm$ 0.076 \\
\vspace{0.1cm}
QNMC (IE) Resc 4 & {\bf 0.314 $\pm$ 0.081} \\
\vspace{0.1cm}
NMC & 0.495 $\pm$ 0.085 \\
\vspace{0.1cm}
LDA & 0.393 $\pm$ 0.082 \\
\vspace{0.1cm}
QDA & 0.568 $\pm$ 0.119 
\\
%

\bottomrule
\end{tabular}
\end{center}
\caption{Average classification error for NMC, QNMC (with difference encodings and different rescaling), LDA and QDA classifiers (in $\%$) over $50$ runs with related standard deviations.}
\label{Tab}
}
\end{table}

In Table \ref{Tab} we present the statistical results that allow to compare the performances - in terms of classification error - of the three standard classifiers described above with the two different variants of the quantum-inspired classifier introduced above. In detail, for each classifier we have evaluated the total error (with the respective standard deviation) obtained by running 50 times the algorithm for each different choise of rescaling (each of them in accord with the different survival degree of the features of the dataset). The standard classifiers we have considered are the NMC, the LDA and the QDA. On the other hand, the proposed quantum-inspired classifiers are the QNMC obtained by the stereographic encoding and the QNMC obtained by the informative encoding. In particular, in this second case three different rescalings have been taken into account.

As shown in Table \ref{Tab}, the QNMC provides in general a meaningful improvement of the accuracy in the classification process with respect to all the three standard classifiers that have been considered and, interestingly enough, the values of the accuracy obtained for the third class are remarkable. In particular, the QNMC based on the informative encoding exhibits better performance than the NMC (about $12\%$) and the QDA (where the difference is very high, about $24\%$). On the other hand, this version of the QNMC exhibits performance similar to the LDA (the difference is about $2\%$): since the LDA is the classifier which takes into account the class distribution by means of the covariance matrix (\emph{i.e.}, we can say it is more “informative”), this result suggests that this specific version of the QNMC is sensitive to the dataset distribution and, consequently, it gives a more accurated classification with respect to the NMC, which does not take into account the data distribution.\\
\noindent Let us note that the “stereographic” QNMC provides a classification accuracy worse than the “informative” QNMC (about $8\%$). As a consequence, this is a remarkable result because it suggests that: \emph{i)} keeping information about the original real vector norm during the encoding process is crucial in order to get a more performing model; \emph{ii)} the choice of the specific encoding is fundamental and strongly affects the final pattern classification.\\
\noindent The final result we discuss concerns the use of the informative encoding together with different rescaling parameters for different features (accordingly with the real different incidence of these features on the probability of survival). In particular, we have rescaled the feature columns “Follow Up Time (days)”, “Oxygen saturation $\%$” and “Heart rate” first by a rescaling parameter equal to $0.1$ (“QNMC (IE) Resc 1”), after by a rescaling factor equal to $10$ (“QNMC (IE) Resc 2”) and finally by a rescaling factor equal to $20$ (“QNMC (IE) Resc 3”). In this regard, we can observe a further improvement in terms of accuracy, up to get a classification error equal to $0.33$. The most interesting result is obtained by concurrently rescaling the feature columns “HRCT Pattern”, “Smoking”, “Smoking Status” by a parameter equal to $600$ and the columns “Sex” and “Oxygen saturation $\%$” by a parameter equal to $10$. In this case, we reach a classification error equal to $31 \%$ (“QNMC (IE) Resc 4”), which is much lower than the NMC classification error (indeed, they differ by approximately $20 \%$). \\
Let us remark that in the proposed approach, which consists of rescaling the feature columns by a real parameter in order to reach some computational benefits, we have adopted a systematic empirical procedure in order to get favorable rescaling parameters.  Nevertheless, by the preliminary results shown in Table \ref{Tab}, it is possible to note that - in accord with the \emph{a priori} assignment of the incidence of each feature - we obtain advantages in terms of classification performance by multiplying more significant features by a higher rescaling parameter and less significant features by a lower rescaling parameter. Consequently, we can look at the rescaling factor as a “weight” which plays in accord with the relevance of a specific feature column. It suggests, as a future work, a theoretical analysis in order to systematically obtain the more convenient rescaling for each feature of a given dataset.

\noindent We conclude the experimental sections with the following two remarks: 
\begin{enumerate}
\item even if it is possible to establish whether a classifier is “good” or “bad” for a given dataset by the evaluation of some a priori data characteristics, generally it is no possible to establish an \emph{absolute} superiority of a given classifier for any dataset, thanks to the No Free Lunch Theorem \cite{DuHa}. Anyway, the QNMC seems to be particularly convenient when the data distribution is difficult to treat with the standard NMC;
\item clearly, there exist classifiers more sophisticated than the ones we have considered in the present work for this specific dataset. Anyway, the introduced preliminary results are enough to show that our quantum-inspired minimum distance model outperforms not only its natural classical counterpart (\emph{i.e.}, the NMC) but also other more performing minimum distance methods. 
\end{enumerate}
\section{Concluding remarks}

This paper is mostly devoted to show the potentialities to use the standard framework of the quantum mechanics in the context of classification problem related to biomedical problems. In particular, we have shown how for some artificial and real datasets, some kind of quantum-inspired classifier provides a remarkable improvement of the classification accuracy with respect to some standard classifier. In particular, in the second part of the paper we have focused on a  very special dataset obtained by a real biomedical context. 
Obviously, techniques used in biometrics are actually much more sophisticated with respect to the standard classifiers that we have considered in this work; anyway, we think the results provided in this paper as promising in order to establish a new investigation based on the application of the quantum framework on the biometric classification problems.
In particular, our future investigation will be based on three relevant points: \emph{i)} first, the QNMC arises as a kind of \emph{quantum replacement} of the standard NMC. We think that, exploiting the expressive power of the quantum framework, to follow the same strategy in order to make an analogue quantum replacement of some more sophisticated standard classifier should be naturally beneficial; \emph{ii)} as we have remarked in the paper, the choise of the \emph{best} encoding is strongly dataset-dependent. Anyway, this point deserves a further investigation; as an example, it should be important to identify some class of datasets that, because of some its internal property, is better to manage by using some encoding instead of any other; \emph{iii)} finally,  as we can see, the quantum-inspired classification process we have considered is strongly based on the distribution of the patterns. Hence, the role of the distance is crucial. However, the IPF dataset also contains features that are not given by ordered values (such as \emph{Sex} or \emph{Smoking status}). On this basis, should be useful to modify the datatset trying to keep the same reliability in the classification process but only involving features with ordered values. Obviously, all these points require a very interdisciplinary investigation and some partial result will be introduced in a future work.

Finally, we think that even if our investigation is in a preliminary stage, the actual results introduced in the present paper (and in the previously mentioned ones \cite{sergioli2016,S,SaSe,Entropy}) are promising enough to suggest to carry on with this research.

%
%

\begin{acknowledgements}
This work is supported by the Sardinia Region Project “Time-logical evolution of correlated microscopic systems”, CRP 55, LR 7/8/2007.
\end{acknowledgements}

\end{document}